\documentclass[prl,twocolumn,showpacs]{revtex4}
\usepackage{amsmath,amsthm,amssymb,amscd,float}
\usepackage[latin1]{inputenc}
\usepackage{graphicx}
\usepackage[T1]{fontenc}
\usepackage{mathptmx} 
\newcommand{\al}{\alpha}

\newcommand{\de}{\delta}
\newcommand{\ep}{\epsilon}

\newcommand{\om}{\omega}
\newcommand{\si}{\sigma}
\newcommand{\te}{\theta}

\newcommand{\Ga}{\Gamma}

\newcommand{\hK}{\hat{K}}

\newcommand{\RR}{{\mathbb R}}
\newcommand{\es}{{\mathbb{S}^1}}
\newcommand{\ZZ}{{\mathbb Z}}
\newcommand{\cC}{{\mathcal C}}

\newcommand{\ms}{\mspace{1mu}}
\renewcommand{\leq}{\leqslant}

\def\bbuildrel#1_#2^#3{\mathrel{\mathop{\kern0pt #1}\limits_{#2}^{#3}}}
\newcommand{\tends}[1]{\bbuildrel{\hbox to 2em{\rightarrowfill}}_{#1}^{}}
\newcommand{\I}{\mathrm{i}}
\newcommand{\e}{\mathrm{e}}

\newcommand{\dd}{\mathrm{d}}
\newcounter{ex}

\def\AdS{\mathrm{AdS}}
\def\ads{\mathrm{AdS}_3}

\def\pd{\partial}
\def\hphi{\hat\phi}

\def\cO{\mathcal O}

\def\hom{_{\rm{hom}}}
\DeclareMathOperator\Imag{Im}
\def\lot{\rm{l.o.t.}}
\begin{document}
\title{Causality and the conformal boundary of AdS in real-time holography}
\author{Alberto \surname{Enciso}}%
\email{aenciso@icmat.es} \affiliation{Instituto de Ciencias
  Matemáticas, Consejo Superior de Investigaciones Científicas, 28049
  Madrid, Spain} \author{Niky \surname{Kamran}}%
\email{nkamran@math.mcgill.ca} \affiliation{Department of Mathematics
  and Statistics, McGill University, Montréal, Québec, Canada H3A 2K6}
\date{\today}
\begin{abstract}
  We consider the holographic prescription problem in a (Lorentzian)
  AdS background, deriving from first principles the explicit formulas
  that relate the field at infinity with the field in the bulk. In contrast with the previous studies of the ``real-time'' holography problem, our derivation uses purely classical arguments that involve causality, as in the usual treatment of the holographic prescription problem in Wick-rotated spaces of Euclidean signature.  We show that there is a unique
  propagator that preserves causality and see that this provides a
  simple picture of the relationship between the bulk manifold and its
  conformal boundary.
\end{abstract}
\pacs{11.25.Tq, 11.15.Kc, 02.40.Vh}
\maketitle
%

The gravity/gauge theory duality~\cite{Maldacena} is a conjectured duality between gravitational theories in certain background geometries and dual gauge theories (with no gravity) on their lower-dimensional boundary manifolds. This far-reaching set of ideas occupies a central position in theoretical physics, as this duality implies that strongly coupled gauge theories can be analyzed using gravity (thus opening up new perspectives in the subject) and, conversely, that spacetime geometry emerges from gauge-theoretical data. 

Most authors analyze duality for theories in spaces of Euclidean
signature, which are heuristically obtained from the Lorentzian
ones through Wick rotation. While this ``imaginary time'' formulation
certainly suffices for many purposes, there are also good reasons for
developing the ``real time'' theory. In particular, it is well known
that Wick rotation does not allow to faithfully capture the field
dynamics, so a real-time analysis would be necessary, for example, for
studying non-stationary spacetimes or time-dependent phenomena. An
important example thereof arises in the holographic modeling of the
quark-gluon plasma, which has actually motivated a significant portion
of the recent work on real-time holography (cf.~\cite{Son} and references therein).

Our goal in this Letter is to present a first principles prescription
for determining holographically the field in the bulk from a field on
the boundary, and to analyze the implications of this prescription on
the geometry of the problem. Real-time holography has been discussed
in several papers, including~\cite{Lorentzian} and, especially,
\cite{Son2,Skenderis}, where real-time holographic prescriptions were
given. We will characterize all the possible prescriptions (and
therefore all the associated dynamics on the boundary) in terms of the
data for a mixed initial-boundary value problem. The final
formulas we obtain for the propagators are identical to those
obtained in~\cite{Son2,Skenderis}; in particular, the AdS/CFT
dictionary discussed in these references is still valid in the present
context. What is very different, though, is both the picture that
arises of the underlying boundary manifold and the arguments that lead
to the prescription. In particular, and contrary to what happens
in~\cite{Skenderis}, we will consider neither several copies of the
bulk manifold nor Euclidean caps. Among all possible prescriptions for
the bulk-boundary propagator, we will single out the unique one for
which the propagation is causal. As we will see, this prescription
is entirely holographic.

One remarkable feature of the previous explorations of the holographic
prescription in Lorentzian signature is that, even in the simplest
cases, the discussions rely heavily on quantum field theoretic
arguments. This is totally unlike what happens in its counterpart of
Euclidean signature, that is, in the ball model of hyperbolic space,
as considered e.g.\ in~\cite{Witten}. Hence in our discussion we will
strive to make the discussion entirely ``classical''. As we shall see,
this is will also provide a much clearer picture of the role that the
basic physical properties of the field equations (especially
causality) play in the prescription.

As is customary, we shall present the prescription in the simplest case:
a scalar, possibly massive field $\phi$ living in anti-de Sitter space
$\AdS_{d+1}$. For simplicity of notation, we will carry out the
discussion in dimension $2+1$; the arguments extend almost verbatim to the
higher-dimensional case. The metric can be written as
\[
\dd s^2= \frac{-\dd t^2+\dd x^2+\cos^2 x\;\dd\theta^2}{\sin^2 x}\,,
\]
where $\ads$ is covered by the coordinates $(x,\theta,t)$ whose
respectives ranges are the interval $(0,\frac\pi2]$, the circle
$\es:=\RR/2\pi\ms\ZZ$ and the real line. 
(We are using the same coordinates as in~\cite{Wald}, where the reader
can find more detailed descriptions.) There is a coordinate singularity at
the ``origin'' $x=\frac\pi2$, where the spacelike spheres
$\{x\}\times\es$ contract to a point. We will identify the boundary of
$\ads$ with the cylinder  $\{x=0\}$, so that it can be described by
coordinates $(\theta,t)\in\es\times\RR$ and is naturally endowed with the conformal Lorentzian metric $-\dd t^2+\dd\theta^2$. 

Mathematically, mapping a scalar field $\phi_0(\theta,t)$ in the
boundary to a field $\phi(x,\theta,t)$ in the bulk is some kind of
``boundary problem at infinity'' for a wave equation in $\ads$, which
is a rather nonstandard PDE problem that nevertheless arises naturally in the
context of the gravity/gauge theory duality. The equation that governs a scalar field in $\ads$
is $\square\phi=m^2\phi$, where $\square$ is the wave operator and
for simplicity we assume that the mass $m$ is positive. In our coordinates,
this equation reads
\begin{equation}\label{eqx}
\pd_t^2\phi-\pd_x^2\phi-\frac{\pd_x\phi}{\sin x\;\cos x}-\frac{\pd_\theta^2\phi}{\cos^2 x}+\frac{m^2\phi}{\sin^2x}=0\,.
\end{equation}

We will be interested in reasonably well-behaved fields that do not grow exponentially fast in time (in particular, for fixed $x$ they define a tempered distribution), so we can Fourier-transform $\phi$ in $\theta$ and $t$. The resulting field will be denoted by $\hphi(x,k,\om)$, with $k\in\ZZ$ and $\om\in\RR$. In the Fourier picture, Eq.~\eqref{eqx} then reads
\begin{equation}\label{eqk}
\bigg[\pd_x^2+\frac1{\sin x\;\cos x}\pd_x -\frac{k^2}{\cos^2 x}-\frac{m^2}{\sin^2x}+\om^2\bigg]\hphi(x,k,\om)=0\,.
\end{equation}
This is an ODE of hypergeometric type, whose general solution is the linear combination
\begin{equation*}
\hphi(x,k,\om)=\cC(k,\om)\,G(x,k,\om)+\tilde \cC(k,\om)\,\tilde G(x,k,\om)\,,
\end{equation*}
where
\begin{align*}
G(x,k,\om)&:=(\sin x)^{1+\nu}(\cos x)^{\si} \, F\big(\zeta^\om_{v,\si},\zeta^{-\om}_{v,\si},1+\si;\cos^2x\big)\,,\\
\tilde G(x,k,\om)&:=(\sin x)^{1+\nu}(\cos x)^{-\si} \, F\big(\zeta^\om_{v,-\si},\zeta^{-\om}_{v,-\si},1-\si;\cos^2x\big)\,,
\end{align*}
$F(a,b,c;z)$ stands for the Gauss hypergeometric function,
$\zeta^\om_{\nu,\si}:=(\nu+\si+\om+1)/2$ and  we use the notation
$\nu:=(m^2+1)^{1/2}$ and  $\si:=(k^2+\tfrac14)^{1/2}$.
Since the solution $\tilde G$ does not lead to a well-behaved function at
the point $x=\frac\pi2$, it follows that we must have
\begin{equation}\label{G}
\hphi(x,k,\om)=\cC(k,\om)\,G(x,k,\om)\,,
\end{equation}
where the ``constant'' $\cC(k,\om)$ is, in general, a distribution in the variables~$(k,\om)$. 

Our goal now is to ascertain to what extent the coefficients
$\cC(k,\om)$ in Eq.~\eqref{G} are determined by the condition that
$\phi$ tends to $\phi_0$, in an appropriate sense, at infinity. To
this end, let us consider the behavior of the function $G(x,k,\om)$ as
$x\to0$. For simplicity, we will make the technical assumption that
$\nu$ is not an integer, which is true for almost all values of the
mass $m$ and implies that $\nu>1$; the case of integer $\nu$ can be
dealt with similarly. In this case, by the well-known formula~\cite[15.3.6]{AS70} the function $G$ can be decomposed as
\begin{equation}\label{Gpm}
G(x,k,\om)=\al_-(k,\om)\,G_-(x,k,\om)+\al_+(k,\om)\, G_+(x,k,\om)\,,
\end{equation}
where the functions
\begin{multline*}
G_\pm(x,k,\om):=(\sin x)^{1\pm\nu}(\cos x)^\si\times\\ F\big(\zeta^\om_{\pm\nu,\si},\zeta^{-\om}_{\pm\nu,\si},\tfrac12\pm(\nu-\tfrac12);\sin^2 x\big)
\end{multline*}
respectively have the asymptotics $x^{1\pm\nu}+\cO(x^{2\pm\nu})$ as
$x\to 0$, and the coefficients
\[
\al_\pm(k,\om):=\frac{\Ga(1+\si)\,\Ga(\mp\nu)}{\Ga(\zeta^{\om}_{\mp\nu,\si})\,\Ga(\zeta^{-\om}_{\mp\nu,\si})}
\]
are well-behaved functions of $k$ and $\om$.  

Since $G_-$ dominates $G_+$ at infinity, whenever the
coefficient $\al_-(k,\om)$ is nonzero, $G(x,k,\om)$ then behaves as a
multiple of $x^{1-\nu}$ for small $x$, so it is apparent that
the assertion that $\phi$ tends to $\phi_0$ at infinity should make
precise the idea that $\phi(x,\theta,t)$ behaves as
$x^{1-\nu}\,\phi_0(\theta,t)$ as $x\to0$. A convenient way of making
sense out of this intuition~\cite{Jaming} is to {\em define}\/ that
the function $\phi(x,\theta,t)$ has (generally, a distribution)
$\phi_0(\theta,t)$ as boundary value if for any test function
$\psi(\theta,t)$
\[
\lim_{x\to0^+}x^{\nu-1}\int \phi(x,\theta,t)\,\psi(\theta,t)\,\dd\theta\,\dd t=\int \phi_0(\theta,t)\,\psi(\theta,t)\,\dd\theta\,\dd t\,,
\]
where henceforth all integrals (or sums) are to be understood in the
sense of distributions. Equivalently, we could have expressed this condition in terms of the Fourier representations of $\phi,\phi_0$ and $\psi$, so from Eqs.~\eqref{G} and~\eqref{Gpm} and from the behavior of $G_\pm(x,k,\om)$ as $x\to0$ we infer that the Fourier transforms of the field in the bulk $\phi$ and the boundary field $\phi_0$ are related via the distributional equation
\begin{equation}\label{PV}
\al_-(k,\om)\, \cC(k,\om)=\hphi_0(k,\om)\,.
\end{equation}

Our task is now to compute all solutions to the above equation, which in
turn encode all possible prescriptions for the bulk field $\phi$ compatible with the
boundary field $\phi_0$. We will start by computing a particular
solution to the inhomogeneous equation~\eqref{PV}. The basic idea is
that $\cC(k,\om)$ must be some regularization of $\hphi_0(k,\om)/\al_-(k,\om)$, so
that by Eq.~\eqref{G} the relation between $\phi$ and $\phi_0$ would
be essentially
\begin{equation}
\hphi(x,k,\om)=  H(x,k,\om)\,\phi_0(k,\om)\,,
\end{equation}
with
\begin{equation}\label{H}
 H(x,k,\om):= G(x,k,\om)/\al_-(k,\om)\,.
\end{equation}

A problem with the above formula is that $\al_-(k,\om)$ vanishes at
the points $\om=\pm\om_{kj}$, with
\[
\om_{kj}:=1+\si+\nu+2j\,,\quad j=0,1,\dots\,,
\]
so that the function $H(x,k,\om)$ has poles at these points. It can be
checked that these poles are simple. Following the treatment of the
wave equation in Minkowski space~\cite{Itzykson}, it is natural to
take a small positive parameter $\ep$ and try to define a propagator
as
\begin{equation}\label{hK}
\hK(x,k,\om)=\lim_{\ep\to0^+}  H(x,k,\om-\I\ep)\,.
\end{equation}
This is analogous to the retarded propagator in Minkowski space. In
fact, if one formally applies the well-known reasoning using complex
integration and the
residue theorem as in~\cite[Sect.~1.3.1]{Itzykson},
one formally derives that the inverse Fourier transform of
$\hK(x,k,\om)$ with respect to time, which we denote by $K(x,k,t)$, is identically
zero for $t<0$. Hence from~\eqref{hK} one derives, again formally, that
\begin{equation}\label{causal}
\phi(x,\te,t)=\sum_{k=-\infty}^\infty \int_{-\infty}^t
K(x,k,t-t')\,\e^{\I k(\te-\te')}\,\phi_0(\te',t')\,\frac{\dd \te'\, \dd t'}{2\pi}\,.
\end{equation}
This formula, once it is proved to be valid, will make it manifest that this particular prescription for the bulk-boundary propagator is causal.

Unfortunately, making sense out of this formula is considerably harder
than it is for its counterpart in Minkowski space. In Minkowski space, the issue is settled by introducing an additional exponentially decaying factor of the form
$\e^{-\eta|t|}$ in the integrand to obtain converging Fourier integrals and then taking
the limit $\eta\to0^+$. This is not as straightforward to implement in anti-de Sitter space
because one does not know a priori the behavior of the function
$H(x,k,\om+\I \ep)$ for large values of $\om$ (so we do not even know
if its inverse Fourier transform is well defined as a distribution!)
or for large negative values of $\ep$ (which is crucial in the contour
integration argument~\cite{Itzykson}). It should be noted that these
problems arise when considering $ H(x,k,\om-\I\ep)$ for any fixed
$\ep$, and not just only in the limit. For these reasons, we will sketch at the end of this Letter
an argument showing that the limit~\eqref{hK} exists and that the formula~\eqref{causal} is therefore rigorously justified. 

We now parametrize all the different possible prescriptions for the
field $\phi$. According to Eq.~\eqref{G}, the difference between the
constant $\cC$ characterizing this solution and the
prescription~\eqref{causal} (which corresponds to a particular
solution of this equation) will be characterized by a solution $\cC\hom(k,\om)$ to the
homogeneous equation $\al_-\cC\hom=0$. A solution of this equation
must be supported on the zero set of the function $\al_-(k,\om)$, so
it can be readily checked using~\cite[Th.~V.11]{Simon} that $\cC\hom$
must be a sum of delta functions:
\[
\cC\hom(k,\om)=\sum_{-\infty}^\infty\big( a_{kj}^+\,\de(\om-\om_{kj})+a_{kj}^-\,\de(\om+\om_{kj})\big)\,.
\]
Here $a_{kj}^\pm$ are complex constants. Using Eq.~\eqref{G} and
the fact that $G(x,k,\om)=G(x,k,-\om)$ we
can write the bulk field corresponding to $\cC\hom$ as
\begin{equation}\label{hom}
\phi\hom(x,\te,t)=\sum_{k=-\infty}^\infty \big(a_{kj}^+\,\e^{\I\om_{kj}t}+a_{kj}^-\,\e^{-\I\om_{kj}t}\big)\, \e^{\I
  k\te} G(x,k,\om_{kj})\,.
\end{equation}
The content of Eq.~\eqref{G} is that any prescription for the bulk
field $\phi$ differs from the particular solution~\eqref{causal} by a
function $\phi\hom$ of the form~\eqref{hom} and, conversely, any such
linear combination is a valid prescription for the field $\phi$.

We can now give an interpretation for the complex coefficients
$a_{kj}^\pm$ parametrizing all the possible prescriptions through the
following basic idea. Let us assume, for the sake of simplicity, that
the boundary field $\phi_0$ is smooth and has very good decay
properties; more concretely, let us assume that it is compactly
supported in time and that its angular Fourier transform only has a
finite number of frequencies (functions of this type are dense e.g.\
in the space of $L^2$ functions of $\te$ and $t$). Hence we can take a
large negative time $T$ such that $\phi_0(\te,t)$ is zero for all
$t\leq T$. The particular bulk field given by~\eqref{causal} is then
obviously zero at time $T$, so that the value at time $T$ of any bulk
field $\phi$ associated with this boundary condition will coincide
with that of the homogeneous field~\eqref{hom}. Now, by separation of
variables and Sturm--Liouville theory~\cite{Simon}, the functions
$\psi_{kj}(x,\te):=\e^{ik\te}\, G(x,k,\om_{kj})$ form an orthogonal
basis of the spatial $L^2$ space. Hence one can expand the value of
the field $\phi$ and its time derivative $\pd_t\phi$ at time $T$ in
this basis and find out that the coefficients in this expansion are in
a one-to-one correspondence with the coefficients $a_{kj}^\pm$. More
concretely, if we denote respectively by $A_{kj}$ and $B_{kj}$ the
components of the field and its velocity at time $T$ in the basis
$\psi_{kj}$, one readily finds that they determine the coefficients
$a_{kj}^\pm$ characterizing the holographic prescription through the
relation
\begin{equation}\label{apm}
a_{kj}^\pm=\frac{\e^{\mp\I\om_{kj}T}}2 \bigg(A_{kj}\pm \frac{B_{kj}}{\I\om_{kj}}\bigg)\,.
\end{equation}

Eq.~\eqref{apm} simply asserts that any choice of bulk-boundary
propagator is completely specified by a choice of ``initial
conditions'', that is, of the choice of the value of the field and its
time derivative at any fixed time $T$. From this it follows that the
only choice for the bulk field $\phi$ that is causal corresponds to
$\phi\hom=0$, i.e., is given by~\eqref{causal}. This is because,
when the boundary field is compactly supported,
causality demands that $(\phi,\pd_t\phi)$ should vanish for sufficiently large
negative times $T$, essentially because the boundary condition has not yet
been ``switched on''. But by Eq.~\eqref{apm}  this is only possible
when all the coefficients $a_{kj}^\pm$ are zero, which yields $\phi\hom=0$, and proves the
uniqueness of the causal propagator.

Some remarks are now in order. The main result in this Letter is the
derivation of the formula~\eqref{causal} for a holographic
prescription in an $\AdS$ background (with its Lorentzian signature),
which we have obtained from first principles without resorting to any
QFT methods. This is considerably harder than in the Euclidean
case~\cite{Witten}, since one needs to deal with the nonuniqueness of
solutions to the prescription problem, an issue that arises from the absence of
ellipticity and manifests itself in the existence of multiple
solutions to Eq.~\eqref{PV}. All possible solutions to the
prescription problem in AdS geometry have been characterized in this Letter in terms of 
data corresponding to a mixed boundary-initial value problem.

It should be stressed that, from the point of view adopted in this
Letter, the causal prescription~\eqref{causal} is fully holographic in
the sense that, as it corresponds to choosing zero Cauchy data below
the support of the boundary field (or, to put it differently, at
``$t=-\infty$''), it describes the case where the whole bulk field is
radiated from the boundary. In our approach, it is the only
prescription for the propagator with this property.

It is worth pointing out that nearly all our computations so far are purely
distributional (the only but important exception is the justification
of the validity of the formula~\eqref{causal}, which we shall present
below). We also note that the use of energy estimates will lead to existence
results in more general spacetimes but not to the explicit
representation~\eqref{causal} for the causal propagator (and actually
for any other propagator~\eqref{hom}). We will address this problem 
in a future work. The regularity conditions we have imposed on the boundary field are
inessential: one can easily deal with more general conditions through
a standard approximation argument.

We should remark at this point the picture of the conformal boundary of $\AdS$ that
emerges from our treatment is rather different from what has been
previously suggested. The most comprehensive treatment to date of the
real-time holographic prescription problem is surely that of Skenderis
and van Rees~\cite{Skenderis}. We have seen that our boundary manifold
is a cylinder, which is consistent with the fact that $\AdS$ (our bulk
manifold) is topologically a solid cylinder. Furthermore, the boundary
manifold inherits a conformal Lorentzian metric, so its associated
gauge theories are obviously expected (and can be shown) to be
governed by wave-type equations, as befits a Lorentzian background. In
Skenderis and van Rees's picture, the boundary manifold consists of
several components, the actual number of which depends on the choice
of the bulk-to-boundary propagator. It involves several copies of the Lorentzian
cylinder that appears in our picture plus several spherical caps,
which do not carry a Lorentzian metric but a Euclidean one. The bulk
manifold that is holographically filled by the datum on the boundary
in their picture consists of several copies of $\AdS$ and the
(Euclidean) hyperbolic half-space associated with each cap. The
holographic prescription is finally obtained by an ingenious argument
which posits that the fields in the different components of the bulk
should satisfy certain matching
conditions. 

To conclude, let us sketch the argument that allows us to prove the
existence of the causal propagator of kernel $K(x,\te,t)$. The first
thing one has to prove is that the inverse time Fourier transform of
$ H(x,k,\om-\I\ep)$ exists as a distribution, when $\ep$ is a
positive number. For this it suffices to show that the absolute value
of $ H$ is bounded by a power of $\om$, i.e., that is does not grow
exponentially fast for large $|\om|$ (this shows that $ H$ is a tempered
distribution in $\om$, which is the most convenient setting for
Fourier transforms~\cite{Simon}). For this, let us start by recalling (see~\eqref{H})
that $ H$ is of the form
\[
\frac{ H(x,k,\om)}{A_1(x,k)}=\Ga\Big(\frac{a+\om}2\Big) \,
\Ga\Big(\frac{a-\om}2\Big)\, F\Big(\frac{b+\om}2,\frac{b-\om}2,c;\cos^2x\Big)
\]
where $A_1(x,k),a,b$ and $c$ are quantitues that depend on $k$ but not on
$\om$ (they can be easily read off~\eqref{G} and
\eqref{PV}), and where we considering any complex value of
$\om$. Using the reflection formula~\cite[6.1.17]{AS70} and the
asymptotics of the Gamma function~\cite[6.1.47]{AS70} we readily derive that
\begin{equation*}
\Ga\Big(\frac{a+\om}2\Big) 
\Ga\Big(\frac{a-\om}2\Big)=\frac{C\Ga(\frac{\om+a}2) \Ga(\frac{\om-a}2)^{-1}}{(\om-a)\sin\frac{\pi(\om-a)}2}=\frac{C'\om^{a-1}+\lot}{\sin\frac{\pi(\om-a)}2}
\end{equation*}
when $|\om|$ is large for a nonzero, $k$-dependent constant $C'$. For
the hypergeometric function, there is a uniform
asymptotic formula due to Jones~\cite{Jones} which ensures that
\[
F\Big(\frac{b+\om}2,\frac{b-\om}2,c;\cos^2x\Big)= \frac{A_2(x,k) \,
 J_{c-1}\big(\om(\tfrac\pi2-x)\big)}{\om^{c-1}}+\lot
\]
Here $A_2(x,k)$ is independent of $\om$ and $J_{c-1}$ denotes the Bessel
function, which admits the asymptotic expansion~\cite[9.2.1]{AS70}
\[
J_{c-1}\big(\om(\tfrac\pi2-x)\big)=\frac{A_3(x,k)}{\om^{1/2}}\,\sin \big(\om(\tfrac\pi2-x)+A_3(x,k)\big)+\lot,
\]
where $A_3(x,k)$ is independent of $\om$. Put together, these formulas imply that for any nonzero
$\ep$ and large real values of $|\om|$ one has the power-type bound $| H(x,k,\om-\I\ep)|\leq
A_4(x,k,\ep)\,|\om|^{b-c-\frac12}$, which guarantees that
these functions have a well defined Fourier transform. Here $A_{4}$ is
independent of $\om$ and blows up only for $\ep\to0$.

In order to use the contour integration argument to show that the
inverse time
Fourier transform of $ H(x,k,\om-\I\ep)$ (for a fixed positive $\ep$) is zero for $t<0$, one needs to
show that one can close the contour~\cite{Itzykson}, which amounts to showing that for
$t<0$ and frequencies $\om$ with very large and negative imaginary part, the
contribution to the integral becomes negligible. Since $|\sin
z|=\frac12\e^{|\Imag z|}+\lot$, this follows from the above asymptotic
formulas without much effort. The fact the distributional limit~\eqref{hK} exists
also follows from the asymptotics, as $ H(x,k,\om)$ can the be
decomposed as a regular polynomially bounded function plus a
sum of terms of the form $c_{kj}(\om\pm\om_{kj}-\I\ep)^{-1}$ admitting simple poles, with
polynomially bounded coefficients, that can be dealt with
using a standard argument~\cite[Exercise~V.22]{Simon}.

The authors are indebted to Robert Brandenberger, Keshav Dasgupta,
Alex Maloney, Kostas Skenderis, Balt van Rees and Yi Wang for valuable
comments. A.E.\ is financially supported by the Ramón y Cajal program of the
Spanish Ministry of Science and thanks McGill University for hospitality
and support. A.E.'s research is supported in part by the MICINN and
UCM--Banco Santander under grants no.~FIS2011-22566
and~GR58/08-910556. The research of N.K.\ is supported by NSERC grant
RGPIN 105490-2011.

%

\end{document}